\documentclass[aps,prl, amsmath,amssymb,reprint,superscriptaddress,showpacs]{revtex4-1}
\usepackage{bm}
\usepackage{graphicx}
\usepackage{float}
\usepackage{color,soul}
\usepackage[dvipsnames]{xcolor}
\usepackage[hyperindex,colorlinks=true, linkcolor=blue, citecolor=blue, urlcolor=blue, breaklinks=true]{hyperref}
\newcommand{\ket}[1]{|#1\rangle}             
\newcommand{\braket}[2]{\langle#1|#2\rangle} 

\newcommand{\cu}[1]{\left\{ {#1} \right\}}

\begin{document}

\title{Quantum State Discrimination Using the Minimum Average Number of Copies}


\author{Sergei Slussarenko}
\affiliation{Centre for Quantum Dynamics and Centre for Quantum Computation and Communication Technology, Griffith University, Brisbane, Queensland 4111, Australia}

\author{Morgan M. Weston}
\affiliation{Centre for Quantum Dynamics and Centre for Quantum Computation and Communication Technology, Griffith University, Brisbane, Queensland 4111, Australia}

\author{Jun-Gang Li}
\affiliation{Centre for Quantum Dynamics and Centre for Quantum Computation and Communication Technology, Griffith University, Brisbane, Queensland 4111, Australia}
\affiliation{School of Physics, Beijing Institute of Technology, Beijing 100081, China}

\author{Nicholas Campbell}
\affiliation{Centre for Quantum Dynamics and Centre for Quantum Computation and Communication Technology, Griffith University, Brisbane, Queensland 4111, Australia}

\author{Howard M. Wiseman}
\email{h.wiseman@griffith.edu.au}
\affiliation{Centre for Quantum Dynamics and Centre for Quantum Computation and Communication Technology, Griffith University, Brisbane, Queensland 4111, Australia}

\author{Geoff J. Pryde}
\email{g.pryde@griffith.edu.au}
\affiliation{Centre for Quantum Dynamics and Centre for Quantum Computation and Communication Technology, Griffith University, Brisbane, Queensland 4111, Australia}


\begin{abstract}
In the task of discriminating between nonorthogonal quantum states from multiple copies, the key parameters are the error probability and the resources (number of copies) used. 
Previous studies have considered the task of minimizing the average error probability for fixed resources.  Here we introduce a new state discrimination task: minimizing the average resources for a fixed admissible error probability. 
We show that this new task is not performed optimally by previously known strategies, and derive and experimentally test a detection scheme that performs better.

\end{abstract}


\maketitle

Quantum measurement and control science provides a toolkit for implementing quantum information protocols, overcoming noisy operation, minimizing the use of costly quantum resources, and finding optimal solutions for manipulating and reading out information encoded in quantum states~\cite{book_wiseman_2009}. Understanding how different quantum measurement and control strategies apply in different quantum information tasks is central to realizing the full potential of quantum approaches~\cite{higgins07,prevedel07,gillett10,xiang11,reed12,yonezawa12,liu16}. The importance of finding different quantum measurement and control solutions to different problems is revealed in the task of quantum state discrimination, a primitive for quantum information science and technology. 

Orthogonal quantum states can be distinguished perfectly by quantum measurements. By contrast, when a quantum system is prepared in one of two known nonorthogonal states, no measurement is able to deterministically tell which state the system is in. This characteristic feature of quantum mechanics is one of the key elements for secure quantum communication~\cite{bennett92d,rev_gisin02,vanenk02} and the  corresponding task of nonorthogonal states discrimination has many applications in quantum information science~\cite{book_wiseman_2009,rev_barnett09,brunner13,rev_bae15}. Two types of imperfect state discrimination tasks have been considered to date. Unambiguous state discrimination performs a measurement that is always correct when it provides an answer, but also has a nonzero chance of providing an inconclusive result~\cite{ivanovich87,becerra13}.  A complementary approach, minimum-error discrimination, is one that always provides an answer, but with a nonzero probability of making a mistake~\cite{book_helstrom_1976}.

If $n>1$ copies of the same state are available, measurement strategies within minimum error and unambiguous discrimination tasks use all the available copies in order to minimize the probability of having an incorrect or inconclusive result, respectively. Strategies generally fall into two categories: collective measurements, where a single measurement is performed on all the copies; and local (or  individual) measurements, where each copy of the state is measured separately~\cite{acin05}. It has been shown theoretically that collective measurements generally outperform local ones, providing the smallest
probability of an error given $n$ copies of the state~\cite{peres91}, but such measurements are hard to implement experimentally, especially for large $n$.  Recently, it was demonstrated that, for pure states, local adaptive measurements---where the next measurement to be performed depends on the outcome of the measurement on the previous copy---can perform minimum error discrimination with exactly the same precision as a collective strategy, for any $n$~\cite{brody96,acin05,higgins09,higgins11}. 

While previously known tasks are defined by their goal of minimizing the average error, a more important goal in many quantum information contexts is minimizing the average resources required while keeping errors below a given bound. This resource-centered consideration is at the heart of fault tolerance for quantum computing~\cite{knill98,aharonov97,bennett00}, has been applied to quantum control strategies for state purification~\cite{wiseman06,wiseman08} and readout~\cite{combes08,combes10}.
The idea is that once a required threshold is reached, it is appropriate to allocate remaining resources elsewhere. This leads us to define a new multiple-copy state discrimination task---\textit{guaranteed bounded error} discrimination---where the goal is to reduce discrimination error below a fixed bound, while minimizing the average number of copies required. Intuitively, one might assume that the strategies that are optimal for multiple-copy minimum error discrimination would be also optimal for the guaranteed bounded error one.

In this Letter,  we experimentally apply two local non-adaptive strategies, previously considered for the minimum error discrimination task, to the guaranteed bounded error discrimination problem for two-state discrimination, also with pure states. We then derive and experimentally demonstrate a new strategy, designed specifically for this new task, that is local and non-adaptive. We find experimentally that our new measurement strategy for guaranteed bounded error discrimination performs at least as well as the other local non-adaptive strategies. Surprisingly, it almost always performs better than the theoretical performance of the optimal strategy for the minimum error discrimination task, the adaptive one noted above. In the most interesting regime, where the error is small, our new scheme even {\em scales} better.  
This is unexpected, because the local adaptive strategy does not rely on knowing the number of copies $n$. In the experiment, we investigate the strategies in depth by comparing the true error probabilities for different measurement outcome sets in each strategy, showing why our new approach outperforms the others. 

Without loss of generality, we can say that our quantum system is a qubit known to be prepared in one of the two nonorthogonal pure states $\ket{\psi_j}=\cos\theta\ket{x}-(-1)^j\sin\theta\ket{y}$, with $j=1,2$ and $\theta\in(0,\pi/4)$, for some orthonormal basis $\{\ket{x},\ket{y}\}$. Note that in this work we use angles in Hilbert space, not angles on the Bloch or Poincar\'{e} spheres. The states are prepared with probabilities $q_1$ and $q_2=1-q_1$ and their overlap is $\braket{\psi_1}{\psi_2}=\cos 2\theta$. In other words, the vectors of the two states have $2\theta$ opening angle between them, with the bisector parallel to $\ket{x}$. We consider local projective measurements in an orthonormal basis $\{\ket{\varphi_1},\ket{\varphi_2}\}$, where $\varphi\in[0,\pi/2)$ and 
\begin{equation}
\ket{\varphi_D}=\cos[\varphi-\frac{\pi}{2}(D-1)]\ket{x}+\sin[\varphi-\frac{\pi}{2}(D-1)]\ket{y}. \label{eq:theo_meas}
\end{equation}
The optimal minimum error discrimination single-copy measurement is a Helstrom measurement~\cite{book_helstrom_1976} with $\varphi^{\textrm{H}}(q_1,\theta)=\frac{1}{2}\arctan[(q_1-q_2)^{-1}\tan2\theta]$ and minimum average probability of error $E^\textrm{H}=\frac{1}{2}-\frac{1}{2}(1-4q_1q_2\cos^2 2\theta)^{1/2}$. For the symmetric $q_1=q_2=0.5$ case, $\varphi^{\textrm{H}}(0.5,\theta)=\pi/4$.

\begin{figure}[t]
\includegraphics[width = 0.45\textwidth]{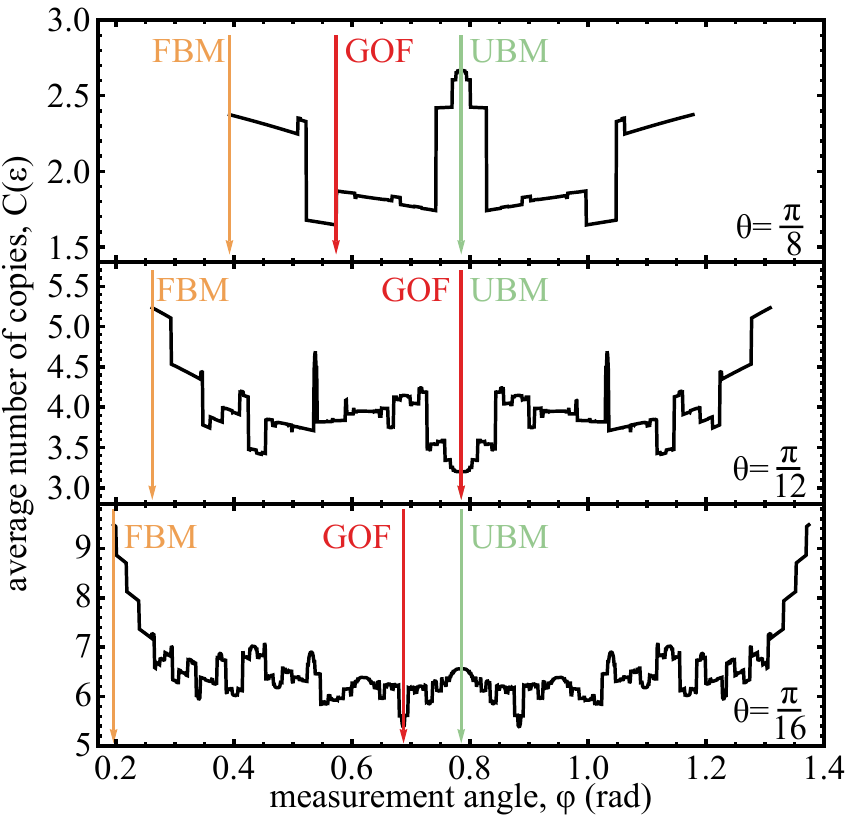}
\caption{\label{fig:theo_strategies}Theoretical prediction of the average number $C(\varepsilon)$ of copies required to perform a state discrimination with a probability of error bounded by $\varepsilon= 0.125$, as a function of measurement angle $\varphi$, for three different symmetric ($q_1=q_2=0.5$) pairs of states with $\theta= \pi/8$, $\pi/12$, and $ \pi/16$. Arrows mark the relevant measurement angle $\varphi$ for each of the three fixed-angle measurement strategies.}
\end{figure}

For multiple copies $n$, we first apply the three previously proposed minimum error discrimination local measurement schemes~\cite{acin05,higgins09} to the guaranteed bounded error discrimination task. The cost function  $C(\varepsilon)$ we consider is the average number of copies of the state required to successfully discriminate the state with the probability of error less than or equal to a preset value of $\varepsilon$. As in Refs.~\cite{higgins09,higgins11} we employ Bayesian processing for our analysis. Plots of theoretically predicted $C(\varepsilon)$ values, for certain parameters, are shown in Fig.~\ref{fig:theo_strategies} for the various schemes which we now explain. 

\emph {1. Fully biased measurements (FBM).} --- In this strategy, a projective measurement with $\varphi=\theta$ is performed individually on each copy of the state. This type of measurement  outputs the guess $\ket{\psi_1}$ for the input state when all the measurement outcomes correspond to $\ket{\varphi_1}$, otherwise, the guess is $\ket{\psi_2}$. The average number of measurements $C^{\textrm{FBM}}(\varepsilon)$  is given by
\begin{equation}
C^{\textrm{FBM}}(\varepsilon)=q_1 n_T+q_2 \frac{1-(\cos 2\theta)^{2n_T}}{(\sin 2\theta)^2}, \label{eq:fbm_cost}
\end{equation}
with $n_T= 
\lceil \cu{\ln(q_1\varepsilon)-\ln[q_2(1-\varepsilon)]}/{2\ln{(\cos2\theta)}}\rceil$,
where $\lceil  \bullet \rceil$ denotes the integer part of $(\bullet + 1)$.

\emph {2. Unbiased measurements (UBM).} --- In this strategy, a single-copy Helstrom measurement is performed individually on each copy of the state. The posterior probability of the quantum system to be found in state $\ket{\psi_1}$, given $m$ of $n$ measurement outcomes are $\ket{\varphi_1}$ is  
\begin{equation}
p_n(\psi_1|m,n)=\left[1+\frac{q_2}{q_1}\left(\frac{1-\sin 2\theta}{1+\sin 2\theta}\right)^{(2m-n)}\right]^{-1}. \label{eq:ubm_Perr}
\end{equation}
This $p_n(\psi_1|m,n)$ is a function of the random variable $R_n=2m-n$. For the $q_1=q_2$ case, 
which we restrict to for our experimental and numerical investigations, the evolution of $R_n$ in discrete `time' $n$ forms a Markov chain for which the transition probabilities can be readily derived. The average number of measurements $C^{\textrm{UBM}}(\varepsilon)$ can be then calculated semianalytically as the average survival time for a random walk on the integers with two absorbing boundaries; see Ref.~\cite{Note1} for details.

\emph {3.Locally optimal local measurements (LOL).} --- This strategy was previously shown to be optimal for multicopy minimum error discrimination~\cite{higgins09}. In this adaptive measurement scheme, an optimal Helstrom measurement with  $\varphi^{H}(q_1)$ is performed on the first copy of the state. The outcome of the measurement is used to update the prior probability $P^{(1)}=q_1$ to posterior probability $P^{(2)}$. This probability is then used to update the  measurement angle $\varphi^{H}(P^{(2)})$ for the next measurement, and the procedure is repeated. For guaranteed bounded error discrimination, the error probability decreases deterministically with the number of copies used, so in this case the average number of measurements required to reach the error bound is exactly the number necessary, the integer
\begin{equation}
C^{\textrm{LOL}}(\varepsilon)=\bigg{\lceil}\frac{\ln(\varepsilon-\varepsilon^2)-\ln(q_1q_2)}{2\ln(\cos 2\theta)}+1\bigg{\rceil}. \label{eq:lol_cost}
\end{equation}

\emph {4. Globally optimal fixed local measurement (GOF).} This is the new strategy which we
now introduce. Like strategies 1 and 2 above, it is a fixed-measurement-angle approach. In this strategy we numerically search for an optimal angle $\varphi^{\textrm{\rm opt}}$ that minimizes the number of copies required for the discrimination of the states with a given angle $\theta$ and a preset error bound $\varepsilon$. The results of numerical simulations are shown in Fig.~\ref{fig:theo_strategies}. As can be seen, $C(\varphi)$ is a not a smooth function and its global minimum does not always corresponds to $\varphi=\pi/4$ (i.e., UBM strategy for the symmetric $q_1=q_2=0.5$ case) or to $\varphi=\theta$ (FBM strategy). The $C(\varepsilon)$ distribution is different for each state angle $\theta$, meaning that the optimal measurement angle $\varphi^{\textrm{opt}}$  is a function of both $\varepsilon$ and  $\theta$.  Further details on the dependence of $\varphi^{\textrm{\rm opt}}$ and  $C(\varepsilon)$ on state angle $\theta$ are presented in Ref.~\cite{Note1}.

\begin{figure}[H]
\includegraphics[width = 0.45\textwidth]{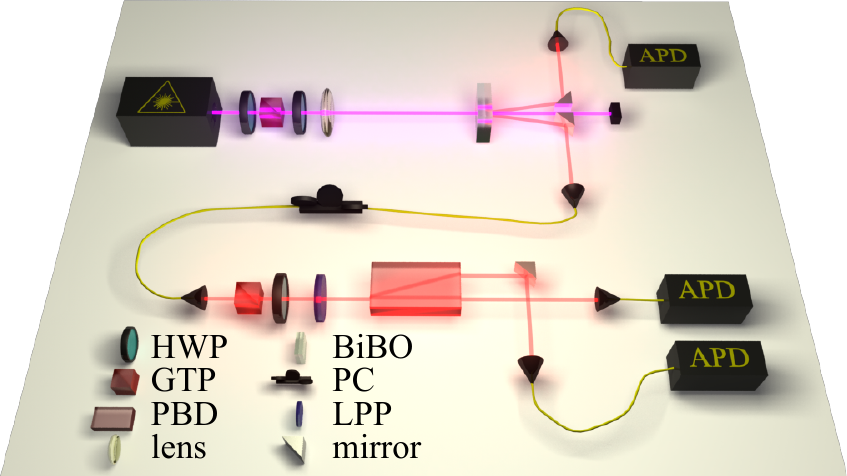}
\caption{\label{fig:exp_setup}Experimental setup. Pairs of single photons were obtained using a $404~\textrm{nm}$ cw diode laser pump and a nonlinear bismuth triborate (BiBO) crystal cut for noncollinear type-I spontaneous parametric down-conversion. The generated photon pairs of $808~\textrm{nm}$ wavelength were spectrally filtered by interference filters ($3~\mathrm{nm}$ FWHM) and collected into optical fibers. One photon from the pair was used as a herald, while the other was sent through the fiber into the state preparation and measurement setup. A combination of a fiber polarization controller (PC), a Glan-Taylor polarizer (GTP), a half-wave plate (HWP), and a rotating linear plate polarizer (LPP) were used to prepare the linear polarization state and polarizing beam displaced (PDB) was used for the high-fidelity projective polarization measurement. APDs are avalanche photodiodes.}
\end{figure}

We tested the applicability of our theoretical predictions by experimentally discriminating nonorthogonal linear polarization states of single photons with $\theta=\pi/12$ and $q_1=q_2=0.5$,  using the experimental setup shown in Fig.~\ref{fig:exp_setup}. The use of a linear plate polarizer (LPVISB050-MP2 from Thorlabs) allowed us to have high quality state preparation  across the entire $\theta\in[0,\pi/2)$. Since the  measurement outcomes are dependent only on the relative angle $\theta-\varphi$ between the state and measurement axes, not their global orientation, the  measurement axes in our experiment were fixed to $\{\ket{x},\ket{y}\}$, while the angle of the prepared polarization state was adjusted to provide the required $\theta-\varphi$ settings. This allowed us to use a fixed polarizing beam displacer as the measurement device, which provided high-contrast-ratio state projection on $\{\ket{x},\ket{y}\}$ basis states (see Ref.~\cite{Note1} for details on state characterization). 

\begin{figure}
\includegraphics[width = 0.45\textwidth]{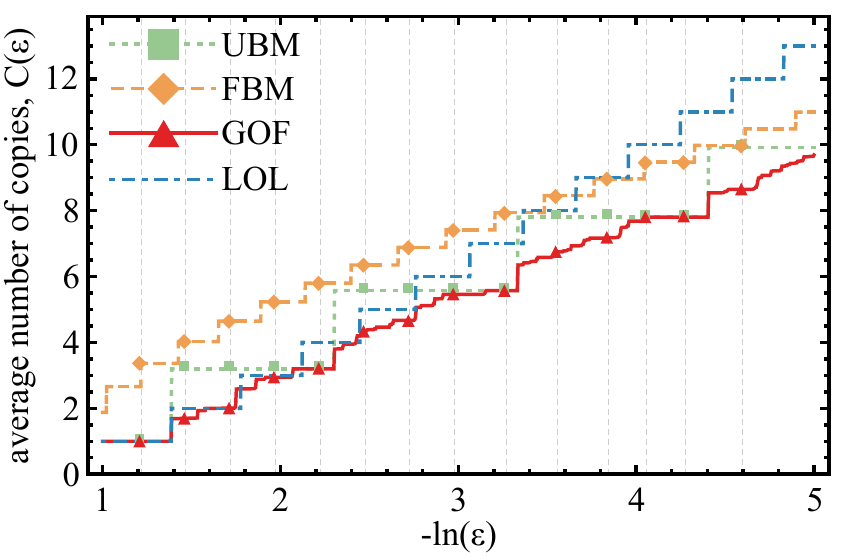}
\caption{\label{fig:exp_strategies}Average number of copies $C(\varepsilon)$ required to perform state discrimination as described in the text, as a function of preset error bound $\varepsilon$, using four different discrimination strategies. Lines represent theoretical predictions for all four discrimination strategies, including the adaptive LOL strategy~\cite{higgins09}. The state angle is $\theta=\pi/12$. Points represent experimental data for fixed-measurement-angle strategies; each point was acquired with $100 000$ independent measurement sets. Error bars, arising from the statistical uncertainty due to the finite number of measurement sets, are smaller than the point size.}
\end{figure}

The experimental  results for the three fixed-angle strategies, together with theoretical predictions for all four strategies outlined above, are shown in Fig.~\ref{fig:exp_strategies}.  Each experimental point was calculated from an independent data run and each GOF measurement was acquired with a different $\varphi^{\rm opt}$ value, optimal for its corresponding $\varepsilon$. Our experimental data match the theory well, and confirm that the GOF strategy performs at least as well as the other fixed-angle strategies, and is usually better than any of the other strategies, for the whole range of preset error values. Furthermore, as suggested  by theoretical curves in Fig.~\ref{fig:exp_strategies}, and shown in Ref.~\cite{Note1}, in the asymptotic limit of small $\varepsilon$, the minimum-error-optimal strategy (adaptive LOL), has a worse scaling than  any of the fixed-angle measurement strategies. The latter appear from the numerics to all have the same scaling, although  the GOF strategy still performs better than the FBM strategy all of the time, and better than the UBM most of the time. This reversal in scaling, with the best strategy for minimizing average error for fixed resources becoming the worst strategy for minimizing average resources for fixed error, was also seen in the task of qubit purification~\cite{wiseman06,wiseman08}, with state purity in place of discrimination error. 

\begin{figure}
\includegraphics[width = 0.45\textwidth]{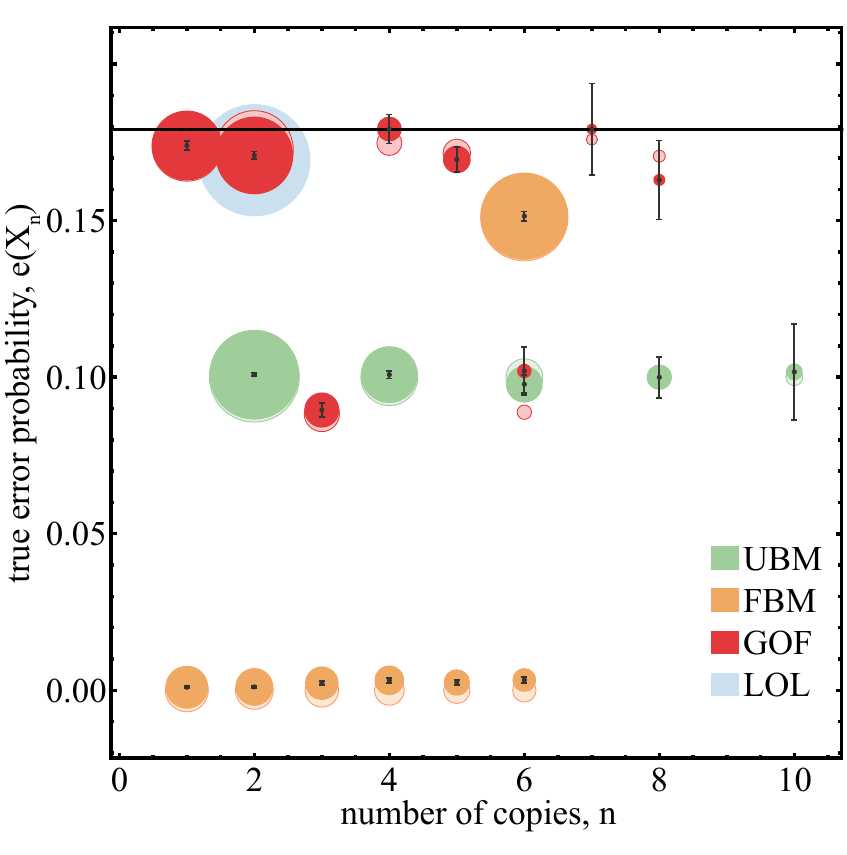}
\caption{\label{fig:exp_trueerrors}Experimentally observed true error probabilities of different successful measurement outcome strings $X_{n}$ for three fixed-angle discrimination strategies, sorted according to the length $n$. The state angle is $\theta=\pi/12$. The target maximum probability of error $\varepsilon= 0.179$ is illustrated by a horizontal line on the plot and corresponds to the vertical dashed line third from the left in Fig.~\ref{fig:exp_strategies}, at $-\ln(0.179)\approx1.716$. Each ball area is proportional to the observed probability $P(X_{n})$ of particular measurement string $X_n$ of length $n$, except for the UBM strategy where different strings are combined (see main text). The vertical position of each ball center corresponds to the observed probability of error $e(X_{n})$ for each string (or set of theoretically equivalent strings for the UBM case). Translucent balls represent theoretical predictions for all four measurement strategies. See main text for details.}
\end{figure}

Insight as to why the fixed-angle GOF strategy performs better than the other fixed-angle strategies can be obtained by analyzing the discrimination measurement procedure in more detail. The outcome of a measurement of a single copy of the state can be represented by the index $D=1,2$, corresponding to projection onto the state $\ket{\varphi_D}$. An outcome of a multiple-copy measurement process on $n$ copies can be then represented by a string  $\chi_{n}=D_1D_2...D_{n}$ of length $n$. 
For a given $n$, only a subset $\mathbb{S}_n$ of 
 these strings will correspond to a discrimination run that is successful (i.e. it allows one to guess the state with an error less than or equal to $\varepsilon$) at the $n$th copy, and is thus terminated then. 
The elements of $\mathbb{S}_n$ will be denoted $X_n$.  
For example, for the FBM measurement strategy, even one $\ket{\varphi_2}$ detection outcome would conclude the measurement procedure with the guess $\ket{\psi_2}$, excluding any combination which is not of form $1_1, 1_2,...,1_{n-1}, 2_{n}$, or $1_1,...,1_{n}$ (of sufficient length to allow one to successfully guess $\ket{\psi_1}$).  

Each $X_{n}$ 
will have its own probability $P(X_{n})$ of being observed as the result of a measurement run.  
Note that we cannot have $X_n  = X'_{n'}D_{n'+1}...D_n$ 
for any $X'_{n'} \in \mathbb{S}_{n'}$, because the measurement would have terminated at copy $n'$. 
Thus, there is no multiple counting of successful measurements, and  
$\sum_{n=1}^{\infty} \sum_{X_n\in \mathbb{S}_n} P(X_{n}) = 1$. 
This allows us to define the cost as 
 $C =  \sum_{n=1}^{\infty} \sum_{X_n\in \mathbb{S}_n} n P(X_{n})$. 
Each  $X_n$ provides a certain \textit{true} probability of discrimination error $e(X_{n})$, which, it is worth noting, may be considerably smaller than $\varepsilon$, especially 
when $\varepsilon$ is just smaller (i.e.,~$-\ln\varepsilon$ is just greater) 
 than a point at which the relevant cost function curve $C(\varepsilon)$ 
changes discontinuously downwards (i.e.,~upwards when plotted versus $-\ln\varepsilon$).  
Such discontinuities occur for all four schemes, not just the LOL 
scheme for which $C(\varepsilon)$ is an integer. 

We have experimentally measured the individual error probabilities $e(X_{n})$ 
and likewise the probabilities $P(X_{n})$ of the successful measurement outcomes, for all three fixed-angle strategies. We choose parameters of $\varepsilon= 0.179$ and state angle $\theta=\pi/12$ to highlight the features of the different schemes. Experimental results, together with corresponding theoretical predictions are shown in Fig.~\ref{fig:exp_trueerrors}. For the UBM measurement strategy, all successful strings have expected true error probability $e(X_{n})=0.1$. These strings $X_n$ are those for which the number of $1$'s exceed (or fall short of) the number of $2$'s by exactly 2, and for which this condition is not satisfied by any substring $X_{n'}$ for $n'<n$. Thus, the UBM ball for a given $n$ (in this case, even numbers) corresponds to the average $e(X_{n})$ and total $P(X_{n})$ for the full  set of observed strings $X_{n} \in \mathbb{S}_n$. Only results for $n\leq10$ are shown, corresponding to a total probability $\sum_{n=1}^{10} \sum_{X_n} P(X_{n}) \approx 99.3\%$. For the FBM measurement strategy, the relevant strings are $\{2,12,112,1112,11112,111112,111111\}$, with the first six corresponding to the outcome $\ket{\varphi_2}$ and expected true probability of error $e(X_{n})=0$, and the remaining one corresponding to the outcome $\ket{\varphi_1}$. For the GOF measurement strategy, the 8 most probable strings, $\{2,11,122,1212,12111,121122,1211212,12112111\}$, are shown, 
corresponding to a total probability of $\approx 99.8\%$. As noted before, the LOL strategy requires 
a deterministic number of copies to perform a discrimination. In this case, 
$C^{\rm LOL}(0.179)=2$, a theoretical prediction illustrated by a single ball at $n=2$ . For  the choice of parameters in this figure, the LOL strategy performs  basically  as well as the GOF one, with $C^{\rm GOF}(0.179)=2.005\pm0.005$, despite having an entirely different distribution of measurement strings $X_{n}$ and individual true error probabilities $e(X_{n})$. 

Our measurements confirm that the true error lies at or beneath the bound for each individual detection string $X_{n}$ that is predicted to lead to a successful discrimination. Additionally, it can be seen that the previously used discrimination strategies (UBM and FBM) provide, in all but one case, true error probabilities far below the bound $\varepsilon$. However, there is a significant chance that one has to probe a large number of copies to reach a successful discrimination---in both cases, a more than $10\%$ chance that six or more copies are needed.  By contrast, the GOF strategy gives, with very high chance a true error probability close to the bound of $\varepsilon= 0.179$. Thus it is able to have a less than $10\%$ chance that more than three copies are needed.

In conclusion, we have studied the problem of multicopy discrimination of two specified nonorthogonal quantum states from the point of view of minimizing resources (number of copies used).  We showed that this guaranteed bounded error discrimination task is not optimized by any of the strategies used in usual multicopy minimum error discrimination. Instead, we introduced a new scheme---the globally optimal fixed measurement scheme---and we demonstrated theoretically and experimentally that it matches or (more usually) outperforms the other options. There is actually one exception to the last statement: for poor discrimination (large error bound), the adaptive scheme that is optimal for minimum-error discrimination sometimes outperforms our new scheme. This raises the interesting possibility that a different adaptive scheme could perform better than our globally optimal fixed measurement scheme in all regimes. This is an issue for future investigation. 

We note that minimizing resources---in this case, the number of copies of the quantum state being used---is often a key consideration in quantum control and quantum information tasks. Moreover, we discovered a reversal in the performance of schemes when swapping the task definition from performance maximization to resource minimization, similar to that previously observed in state purification~\cite{wiseman06,wiseman08}. This suggests that this phenomenon is a generic one, with wide relevance, requiring further work to illuminate it.  

This work was supported by the Australian Research Council Centre of Excellence for Quantum Computation and Communication Technology (Project No. CE110001027). J.G.L. acknowledges financial support from China Scholarship Council. 


\begin{thebibliography}{33}%
\makeatletter
\providecommand \@ifxundefined [1]{%
 \@ifx{#1\undefined}
}%
\providecommand \@ifnum [1]{%
 \ifnum #1\expandafter \@firstoftwo
 \else \expandafter \@secondoftwo
 \fi
}%
\providecommand \@ifx [1]{%
 \ifx #1\expandafter \@firstoftwo
 \else \expandafter \@secondoftwo
 \fi
}%
\providecommand \natexlab [1]{#1}%
\providecommand \enquote  [1]{``#1''}%
\providecommand \bibnamefont  [1]{#1}%
\providecommand \bibfnamefont [1]{#1}%
\providecommand \citenamefont [1]{#1}%
\providecommand \href@noop [0]{\@secondoftwo}%
\providecommand \href [0]{\begingroup \@sanitize@url \@href}%
\providecommand \@href[1]{\@@startlink{#1}\@@href}%
\providecommand \@@href[1]{\endgroup#1\@@endlink}%
\providecommand \@sanitize@url [0]{\catcode `\\12\catcode `\$12\catcode
  `\&12\catcode `\#12\catcode `\^12\catcode `\_12\catcode `\%12\relax}%
\providecommand \@@startlink[1]{}%
\providecommand \@@endlink[0]{}%
\providecommand \url  [0]{\begingroup\@sanitize@url \@url }%
\providecommand \@url [1]{\endgroup\@href {#1}{\urlprefix }}%
\providecommand \urlprefix  [0]{URL }%
\providecommand \Eprint [0]{\href }%
\providecommand \doibase [0]{http://dx.doi.org/}%
\providecommand \selectlanguage [0]{\@gobble}%
\providecommand \bibinfo  [0]{\@secondoftwo}%
\providecommand \bibfield  [0]{\@secondoftwo}%
\providecommand \translation [1]{[#1]}%
\providecommand \BibitemOpen [0]{}%
\providecommand \bibitemStop [0]{}%
\providecommand \bibitemNoStop [0]{.\EOS\space}%
\providecommand \EOS [0]{\spacefactor3000\relax}%
\providecommand \BibitemShut  [1]{\csname bibitem#1\endcsname}%
\let\auto@bib@innerbib\@empty
\bibitem [{\citenamefont {Wiseman}\ and\ \citenamefont
  {Milburn}(2009)}]{book_wiseman_2009}%
  \BibitemOpen
  \bibfield  {author} {\bibinfo {author} {\bibfnamefont {H.~M.}\ \bibnamefont
  {Wiseman}}\ and\ \bibinfo {author} {\bibfnamefont {G.~J.}\ \bibnamefont
  {Milburn}},\ }\href@noop {} {\emph {\bibinfo {title} {Quantum Measurement and
  Control}}}\ (\bibinfo  {publisher} {Cambridge University Press},\ \bibinfo
  {address} {Cambridge, England},\ \bibinfo {year} {2009})\BibitemShut
  {NoStop}%
\bibitem [{\citenamefont {Higgins}\ \emph {et~al.}(2007)\citenamefont
  {Higgins}, \citenamefont {Berry}, \citenamefont {Bartlett}, \citenamefont
  {Wiseman},\ and\ \citenamefont {Pryde}}]{higgins07}%
  \BibitemOpen
  \bibfield  {author} {\bibinfo {author} {\bibfnamefont {B.~L.}\ \bibnamefont
  {Higgins}}, \bibinfo {author} {\bibfnamefont {D.~W.}\ \bibnamefont {Berry}},
  \bibinfo {author} {\bibfnamefont {S.~D.}\ \bibnamefont {Bartlett}}, \bibinfo
  {author} {\bibfnamefont {H.~M.}\ \bibnamefont {Wiseman}}, \ and\ \bibinfo
  {author} {\bibfnamefont {G.~J.}\ \bibnamefont {Pryde}},\ }\href
  {http://dx.doi.org/10.1038/nature06257} {\bibfield  {journal} {\bibinfo
  {journal} {Nature}\ }\textbf {\bibinfo {volume} {450}},\ \bibinfo {pages}
  {393} (\bibinfo {year} {2007})}\BibitemShut {NoStop}%
\bibitem [{\citenamefont {Prevedel}\ \emph {et~al.}(2007)\citenamefont
  {Prevedel}, \citenamefont {Walther}, \citenamefont {Tiefenbacher},
  \citenamefont {Bohi}, \citenamefont {Kaltenbaek}, \citenamefont {Jennewein},\
  and\ \citenamefont {Zeilinger}}]{prevedel07}%
  \BibitemOpen
  \bibfield  {author} {\bibinfo {author} {\bibfnamefont {R.}~\bibnamefont
  {Prevedel}}, \bibinfo {author} {\bibfnamefont {P.}~\bibnamefont {Walther}},
  \bibinfo {author} {\bibfnamefont {F.}~\bibnamefont {Tiefenbacher}}, \bibinfo
  {author} {\bibfnamefont {P.}~\bibnamefont {Bohi}}, \bibinfo {author}
  {\bibfnamefont {R.}~\bibnamefont {Kaltenbaek}}, \bibinfo {author}
  {\bibfnamefont {T.}~\bibnamefont {Jennewein}}, \ and\ \bibinfo {author}
  {\bibfnamefont {A.}~\bibnamefont {Zeilinger}},\ }\href
  {http://dx.doi.org/10.1038/nature05346} {\bibfield  {journal} {\bibinfo
  {journal} {Nature}\ }\textbf {\bibinfo {volume} {445}},\ \bibinfo {pages}
  {65} (\bibinfo {year} {2007})}\BibitemShut {NoStop}%
\bibitem [{\citenamefont {Gillett}\ \emph {et~al.}(2010)\citenamefont
  {Gillett}, \citenamefont {Dalton}, \citenamefont {Lanyon}, \citenamefont
  {Almeida}, \citenamefont {Barbieri}, \citenamefont {Pryde}, \citenamefont
  {O'Brien}, \citenamefont {Resch}, \citenamefont {Bartlett},\ and\
  \citenamefont {White}}]{gillett10}%
  \BibitemOpen
  \bibfield  {author} {\bibinfo {author} {\bibfnamefont {G.~G.}\ \bibnamefont
  {Gillett}}, \bibinfo {author} {\bibfnamefont {R.~B.}\ \bibnamefont {Dalton}},
  \bibinfo {author} {\bibfnamefont {B.~P.}\ \bibnamefont {Lanyon}}, \bibinfo
  {author} {\bibfnamefont {M.~P.}\ \bibnamefont {Almeida}}, \bibinfo {author}
  {\bibfnamefont {M.}~\bibnamefont {Barbieri}}, \bibinfo {author}
  {\bibfnamefont {G.~J.}\ \bibnamefont {Pryde}}, \bibinfo {author}
  {\bibfnamefont {J.~L.}\ \bibnamefont {O'Brien}}, \bibinfo {author}
  {\bibfnamefont {K.~J.}\ \bibnamefont {Resch}}, \bibinfo {author}
  {\bibfnamefont {S.~D.}\ \bibnamefont {Bartlett}}, \ and\ \bibinfo {author}
  {\bibfnamefont {A.~G.}\ \bibnamefont {White}},\ }\href {\doibase
  10.1103/PhysRevLett.104.080503} {\bibfield  {journal} {\bibinfo  {journal}
  {Phys. Rev. Lett.}\ }\textbf {\bibinfo {volume} {104}},\ \bibinfo {pages}
  {080503} (\bibinfo {year} {2010})}\BibitemShut {NoStop}%
\bibitem [{\citenamefont {Xiang}\ \emph {et~al.}(2011)\citenamefont {Xiang},
  \citenamefont {Higgins}, \citenamefont {Berry}, \citenamefont {Wiseman},\
  and\ \citenamefont {Pryde}}]{xiang11}%
  \BibitemOpen
  \bibfield  {author} {\bibinfo {author} {\bibfnamefont {G.~Y.}\ \bibnamefont
  {Xiang}}, \bibinfo {author} {\bibfnamefont {B.~L.}\ \bibnamefont {Higgins}},
  \bibinfo {author} {\bibfnamefont {D.~W.}\ \bibnamefont {Berry}}, \bibinfo
  {author} {\bibfnamefont {H.~M.}\ \bibnamefont {Wiseman}}, \ and\ \bibinfo
  {author} {\bibfnamefont {G.~J.}\ \bibnamefont {Pryde}},\ }\href
  {http://dx.doi.org/10.1038/nphoton.2010.268} {\bibfield  {journal} {\bibinfo
  {journal} {Nat. Photon.}\ }\textbf {\bibinfo {volume} {5}},\ \bibinfo {pages}
  {43} (\bibinfo {year} {2011})}\BibitemShut {NoStop}%
\bibitem [{\citenamefont {Reed}\ \emph {et~al.}(2012)\citenamefont {Reed},
  \citenamefont {DiCarlo}, \citenamefont {Nigg}, \citenamefont {Sun},
  \citenamefont {Frunzio}, \citenamefont {Girvin},\ and\ \citenamefont
  {Schoelkopf}}]{reed12}%
  \BibitemOpen
  \bibfield  {author} {\bibinfo {author} {\bibfnamefont {M.~D.}\ \bibnamefont
  {Reed}}, \bibinfo {author} {\bibfnamefont {L.}~\bibnamefont {DiCarlo}},
  \bibinfo {author} {\bibfnamefont {S.~E.}\ \bibnamefont {Nigg}}, \bibinfo
  {author} {\bibfnamefont {L.}~\bibnamefont {Sun}}, \bibinfo {author}
  {\bibfnamefont {L.}~\bibnamefont {Frunzio}}, \bibinfo {author} {\bibfnamefont
  {S.M.}\ \bibnamefont {Girvin}}, \ and\ \bibinfo {author} {\bibfnamefont
  {R.J.}\ \bibnamefont {Schoelkopf}},\ }\href
  {http://dx.doi.org/10.1038/nature10786} {\bibfield  {journal} {\bibinfo
  {journal} {Nature}\ }\textbf {\bibinfo {volume} {482}},\ \bibinfo {pages}
  {382} (\bibinfo {year} {2012})}\BibitemShut {NoStop}%
\bibitem [{\citenamefont {Yonezawa}\ \emph {et~al.}(2012)\citenamefont
  {Yonezawa}, \citenamefont {Nakane}, \citenamefont {Wheatley}, \citenamefont
  {Iwasawa}, \citenamefont {Takeda}, \citenamefont {Arao}, \citenamefont
  {Ohki}, \citenamefont {Tsumura}, \citenamefont {Berry}, \citenamefont
  {Ralph}, \citenamefont {Wiseman}, \citenamefont {Huntington},\ and\
  \citenamefont {Furusawa}}]{yonezawa12}%
  \BibitemOpen
  \bibfield  {author} {\bibinfo {author} {\bibfnamefont {H.}~\bibnamefont
  {Yonezawa}}, \bibinfo {author} {\bibfnamefont {D.}~\bibnamefont {Nakane}},
  \bibinfo {author} {\bibfnamefont {T.~A.}\ \bibnamefont {Wheatley}}, \bibinfo
  {author} {\bibfnamefont {K.}~\bibnamefont {Iwasawa}}, \bibinfo {author}
  {\bibfnamefont {S.}~\bibnamefont {Takeda}}, \bibinfo {author} {\bibfnamefont
  {H.}~\bibnamefont {Arao}}, \bibinfo {author} {\bibfnamefont {K.}~\bibnamefont
  {Ohki}}, \bibinfo {author} {\bibfnamefont {K.}~\bibnamefont {Tsumura}},
  \bibinfo {author} {\bibfnamefont {D.~W.}\ \bibnamefont {Berry}}, \bibinfo
  {author} {\bibfnamefont {T.~C.}\ \bibnamefont {Ralph}}, \bibinfo {author}
  {\bibfnamefont {H.~M.}\ \bibnamefont {Wiseman}}, \bibinfo {author}
  {\bibfnamefont {E.~H.}\ \bibnamefont {Huntington}}, \ and\ \bibinfo {author}
  {\bibfnamefont {A.}~\bibnamefont {Furusawa}},\ }\href {\doibase
  10.1126/science.1225258} {\bibfield  {journal} {\bibinfo  {journal}
  {Science}\ }\textbf {\bibinfo {volume} {337}},\ \bibinfo {pages} {1514}
  (\bibinfo {year} {2012})}\BibitemShut {NoStop}%
\bibitem [{\citenamefont {Liu}\ \emph {et~al.}(2016)\citenamefont {Liu},
  \citenamefont {Shankar}, \citenamefont {Ofek}, \citenamefont {Hatridge},
  \citenamefont {Narla}, \citenamefont {Sliwa}, \citenamefont {Frunzio},
  \citenamefont {Schoelkopf},\ and\ \citenamefont {Devoret}}]{liu16}%
  \BibitemOpen
  \bibfield  {author} {\bibinfo {author} {\bibfnamefont {Y.}~\bibnamefont
  {Liu}}, \bibinfo {author} {\bibfnamefont {S.}~\bibnamefont {Shankar}},
  \bibinfo {author} {\bibfnamefont {N.}~\bibnamefont {Ofek}}, \bibinfo {author}
  {\bibfnamefont {M.}~\bibnamefont {Hatridge}}, \bibinfo {author}
  {\bibfnamefont {A.}~\bibnamefont {Narla}}, \bibinfo {author} {\bibfnamefont
  {K.~M.}\ \bibnamefont {Sliwa}}, \bibinfo {author} {\bibfnamefont
  {L.}~\bibnamefont {Frunzio}}, \bibinfo {author} {\bibfnamefont {R.~J.}\
  \bibnamefont {Schoelkopf}}, \ and\ \bibinfo {author} {\bibfnamefont {M.~H.}\
  \bibnamefont {Devoret}},\ }\href {\doibase 10.1103/PhysRevX.6.011022}
  {\bibfield  {journal} {\bibinfo  {journal} {Phys. Rev. X}\ }\textbf {\bibinfo
  {volume} {6}},\ \bibinfo {pages} {011022} (\bibinfo {year}
  {2016})}\BibitemShut {NoStop}%
\bibitem [{\citenamefont {Bennett}(1992)}]{bennett92d}%
  \BibitemOpen
  \bibfield  {author} {\bibinfo {author} {\bibfnamefont {C.~H.}\ \bibnamefont
  {Bennett}},\ }\href {\doibase 10.1103/PhysRevLett.68.3121} {\bibfield
  {journal} {\bibinfo  {journal} {Phys. Rev. Lett.}\ }\textbf {\bibinfo
  {volume} {68}},\ \bibinfo {pages} {3121} (\bibinfo {year}
  {1992})}\BibitemShut {NoStop}%
\bibitem [{\citenamefont {Gisin}\ \emph {et~al.}(2002)\citenamefont {Gisin},
  \citenamefont {Ribordy}, \citenamefont {Tittel},\ and\ \citenamefont
  {Zbinden}}]{rev_gisin02}%
  \BibitemOpen
  \bibfield  {author} {\bibinfo {author} {\bibfnamefont {N.}~\bibnamefont
  {Gisin}}, \bibinfo {author} {\bibfnamefont {G.}~\bibnamefont {Ribordy}},
  \bibinfo {author} {\bibfnamefont {W.}~\bibnamefont {Tittel}}, \ and\ \bibinfo
  {author} {\bibfnamefont {H.}~\bibnamefont {Zbinden}},\ }\href {\doibase
  10.1103/RevModPhys.74.145} {\bibfield  {journal} {\bibinfo  {journal} {Rev.
  Mod. Phys.}\ }\textbf {\bibinfo {volume} {74}},\ \bibinfo {pages} {145}
  (\bibinfo {year} {2002})}\BibitemShut {NoStop}%
\bibitem [{\citenamefont {van Enk}(2002)}]{vanenk02}%
  \BibitemOpen
  \bibfield  {author} {\bibinfo {author} {\bibfnamefont {S.~J.}\ \bibnamefont
  {van Enk}},\ }\href {\doibase 10.1103/PhysRevA.66.042313} {\bibfield
  {journal} {\bibinfo  {journal} {Phys. Rev. A}\ }\textbf {\bibinfo {volume}
  {66}},\ \bibinfo {pages} {042313} (\bibinfo {year} {2002})}\BibitemShut
  {NoStop}%
\bibitem [{\citenamefont {Barnett}\ and\ \citenamefont
  {Croke}(2009)}]{rev_barnett09}%
  \BibitemOpen
  \bibfield  {author} {\bibinfo {author} {\bibfnamefont {S.~M.}\ \bibnamefont
  {Barnett}}\ and\ \bibinfo {author} {\bibfnamefont {S.}~\bibnamefont
  {Croke}},\ }\href {\doibase 10.1364/AOP.1.000238} {\bibfield  {journal}
  {\bibinfo  {journal} {Adv. Opt. Photon.}\ }\textbf {\bibinfo {volume} {1}},\
  \bibinfo {pages} {238} (\bibinfo {year} {2009})}\BibitemShut {NoStop}%
\bibitem [{\citenamefont {Brunner}\ \emph {et~al.}(2013)\citenamefont
  {Brunner}, \citenamefont {Navascu\'es},\ and\ \citenamefont
  {V\'ertesi}}]{brunner13}%
  \BibitemOpen
  \bibfield  {author} {\bibinfo {author} {\bibfnamefont {N.}~\bibnamefont
  {Brunner}}, \bibinfo {author} {\bibfnamefont {M.}~\bibnamefont
  {Navascu\'es}}, \ and\ \bibinfo {author} {\bibfnamefont {T.}~\bibnamefont
  {V\'ertesi}},\ }\href {\doibase 10.1103/PhysRevLett.110.150501} {\bibfield
  {journal} {\bibinfo  {journal} {Phys. Rev. Lett.}\ }\textbf {\bibinfo
  {volume} {110}},\ \bibinfo {pages} {150501} (\bibinfo {year}
  {2013})}\BibitemShut {NoStop}%
\bibitem [{\citenamefont {Bae}\ and\ \citenamefont {Kwek}(2015)}]{rev_bae15}%
  \BibitemOpen
  \bibfield  {author} {\bibinfo {author} {\bibfnamefont {J.}~\bibnamefont
  {Bae}}\ and\ \bibinfo {author} {\bibfnamefont {L.-C.}\ \bibnamefont {Kwek}},\
  }\href {http://stacks.iop.org/1751-8121/48/i=8/a=083001} {\bibfield
  {journal} {\bibinfo  {journal} {J. Phys. A}\ }\textbf {\bibinfo
  {volume} {48}},\ \bibinfo {pages} {083001} (\bibinfo {year}
  {2015})}\BibitemShut {NoStop}%
\bibitem [{\citenamefont {Ivanovic}(1987)}]{ivanovich87}%
  \BibitemOpen
  \bibfield  {author} {\bibinfo {author} {\bibfnamefont {I.~D.}\ \bibnamefont
  {Ivanovic}},\ }\href {\doibase 10.1016/0375-9601(87)90222-2} {\bibfield
  {journal} {\bibinfo  {journal} {Phys. Lett. A}\ }\textbf {\bibinfo {volume}
  {123}},\ \bibinfo {pages} {257 } (\bibinfo {year} {1987})}\BibitemShut
  {NoStop}%
\bibitem [{\citenamefont {Becerra}\ \emph {et~al.}(2013)\citenamefont
  {Becerra}, \citenamefont {Fan},\ and\ \citenamefont {Migdall}}]{becerra13}%
  \BibitemOpen
  \bibfield  {author} {\bibinfo {author} {\bibfnamefont {F.~E.}\ \bibnamefont
  {Becerra}}, \bibinfo {author} {\bibfnamefont {J.}~\bibnamefont {Fan}}, \ and\
  \bibinfo {author} {\bibfnamefont {A.}~\bibnamefont {Migdall}},\ }\href
  {\doibase 10.1038/ncomms3028} {\bibfield  {journal} {\bibinfo  {journal}
  {Nat. Commun.}\ }\textbf {\bibinfo {volume} {4}},\  (\bibinfo {year}
  {2013})}\BibitemShut {NoStop}%
\bibitem [{\citenamefont {Helstrom}(1976)}]{book_helstrom_1976}%
  \BibitemOpen
  \bibfield  {author} {\bibinfo {author} {\bibfnamefont {C.~W.}\ \bibnamefont
  {Helstrom}},\ }\href@noop {} {\emph {\bibinfo {title} {Quantum Detection and
 Estimation Theory}}}\ (\bibinfo  {publisher} {Academic Press},\ \bibinfo
  {address} {New York},\ \bibinfo {year} {1976})\BibitemShut {NoStop}%
\bibitem [{\citenamefont {Ac\'{i}n}\ \emph {et~al.}(2005)\citenamefont
  {Ac\'{i}n}, \citenamefont {Bagan}, \citenamefont {Baig}, \citenamefont
  {Masanes},\ and\ \citenamefont {Munoz-Tapia}}]{acin05}%
  \BibitemOpen
  \bibfield  {author} {\bibinfo {author} {\bibfnamefont {A.}~\bibnamefont
  {Ac\'{i}n}}, \bibinfo {author} {\bibfnamefont {E.}~\bibnamefont {Bagan}},
  \bibinfo {author} {\bibfnamefont {M.}~\bibnamefont {Baig}}, \bibinfo {author}
  {\bibfnamefont {L.}~\bibnamefont {Masanes}}, \ and\ \bibinfo {author}
  {\bibfnamefont {R.}~\bibnamefont {Munoz-Tapia}},\ }\href {\doibase
  10.1103/PhysRevA.71.032338} {\bibfield  {journal} {\bibinfo  {journal} {Phys.
  Rev. A}\ }\textbf {\bibinfo {volume} {71}},\ \bibinfo {pages} {032338}
  (\bibinfo {year} {2005})}\BibitemShut {NoStop}%
\bibitem [{\citenamefont {Peres}\ and\ \citenamefont
  {Wootters}(1991)}]{peres91}%
  \BibitemOpen
  \bibfield  {author} {\bibinfo {author} {\bibfnamefont {A.}~\bibnamefont
  {Peres}}\ and\ \bibinfo {author} {\bibfnamefont {W.~K.}\ \bibnamefont
  {Wootters}},\ }\href {\doibase 10.1103/PhysRevLett.66.1119} {\bibfield
  {journal} {\bibinfo  {journal} {Phys. Rev. Lett.}\ }\textbf {\bibinfo
  {volume} {66}},\ \bibinfo {pages} {1119} (\bibinfo {year}
  {1991})}\BibitemShut {NoStop}%
\bibitem [{\citenamefont {Brody}\ and\ \citenamefont
  {Meister}(1996)}]{brody96}%
  \BibitemOpen
  \bibfield  {author} {\bibinfo {author} {\bibfnamefont {D.}~\bibnamefont
  {Brody}}\ and\ \bibinfo {author} {\bibfnamefont {B.}~\bibnamefont
  {Meister}},\ }\href {\doibase 10.1103/PhysRevLett.76.1} {\bibfield  {journal}
  {\bibinfo  {journal} {Phys. Rev. Lett.}\ }\textbf {\bibinfo {volume} {76}},\
  \bibinfo {pages} {1} (\bibinfo {year} {1996})}\BibitemShut {NoStop}%
\bibitem [{\citenamefont {Higgins}\ \emph {et~al.}(2009)\citenamefont
  {Higgins}, \citenamefont {Booth}, \citenamefont {Doherty}, \citenamefont
  {Bartlett}, \citenamefont {Wiseman},\ and\ \citenamefont
  {Pryde}}]{higgins09}%
  \BibitemOpen
  \bibfield  {author} {\bibinfo {author} {\bibfnamefont {B.~L.}\ \bibnamefont
  {Higgins}}, \bibinfo {author} {\bibfnamefont {B.~M.}\ \bibnamefont {Booth}},
  \bibinfo {author} {\bibfnamefont {A.~C.}\ \bibnamefont {Doherty}}, \bibinfo
  {author} {\bibfnamefont {S.~D.}\ \bibnamefont {Bartlett}}, \bibinfo {author}
  {\bibfnamefont {H.~M.}\ \bibnamefont {Wiseman}}, \ and\ \bibinfo {author}
  {\bibfnamefont {G.~J.}\ \bibnamefont {Pryde}},\ }\href {\doibase
  10.1103/PhysRevLett.103.220503} {\bibfield  {journal} {\bibinfo  {journal}
  {Phys. Rev. Lett.}\ }\textbf {\bibinfo {volume} {103}},\ \bibinfo {pages}
  {220503} (\bibinfo {year} {2009})}\BibitemShut {NoStop}%
\bibitem [{\citenamefont {Higgins}\ \emph {et~al.}(2011)\citenamefont
  {Higgins}, \citenamefont {Doherty}, \citenamefont {Bartlett}, \citenamefont
  {Pryde},\ and\ \citenamefont {Wiseman}}]{higgins11}%
  \BibitemOpen
  \bibfield  {author} {\bibinfo {author} {\bibfnamefont {B.~L.}\ \bibnamefont
  {Higgins}}, \bibinfo {author} {\bibfnamefont {A.~C.}\ \bibnamefont
  {Doherty}}, \bibinfo {author} {\bibfnamefont {S.~D.}\ \bibnamefont
  {Bartlett}}, \bibinfo {author} {\bibfnamefont {G.~J.}\ \bibnamefont {Pryde}},
  \ and\ \bibinfo {author} {\bibfnamefont {H.~M.}\ \bibnamefont {Wiseman}},\
  }\href {\doibase 10.1103/PhysRevA.83.052314} {\bibfield  {journal} {\bibinfo
  {journal} {Phys. Rev. A}\ }\textbf {\bibinfo {volume} {83}},\ \bibinfo
  {pages} {052314} (\bibinfo {year} {2011})}\BibitemShut {NoStop}%
\bibitem [{\citenamefont {Knill}\ \emph {et~al.}(1998)\citenamefont {Knill},
  \citenamefont {Laflamme},\ and\ \citenamefont {Zurek}}]{knill98}%
  \BibitemOpen
  \bibfield  {author} {\bibinfo {author} {\bibfnamefont {E.}~\bibnamefont
  {Knill}}, \bibinfo {author} {\bibfnamefont {R.}~\bibnamefont {Laflamme}}, \
  and\ \bibinfo {author} {\bibfnamefont {W.~H.}\ \bibnamefont {Zurek}},\ }\href
  {\doibase 10.1098/rspa.1998.0166} {\bibfield  {journal} {\bibinfo  {journal}
  {Proc. R. Soc. A}\ }\textbf {\bibinfo {volume} {454}},\ \bibinfo {pages}
  {365} (\bibinfo {year} {1998})}\BibitemShut {NoStop}%
\bibitem [{\citenamefont {Aharonov}\ and\ \citenamefont
  {Ben-Or}(1997)}]{aharonov97}%
  \BibitemOpen
  \bibfield  {author} {\bibinfo {author} {\bibfnamefont {D.}~\bibnamefont
  {Aharonov}}\ and\ \bibinfo {author} {\bibfnamefont {M.}~\bibnamefont
  {Ben-Or}},\ }in\ \href {\doibase 10.1145/258533.258579} {\emph {\bibinfo
  {booktitle} {Proceedings of the Twenty-Ninth Annual ACM Symposium on Theory
  of Computing}}},\ \bibinfo {series and number} {STOC '97}\ (\bibinfo
  {publisher} {ACM},\ \bibinfo {address} {New York, NY, USA},\ \bibinfo {year}
  {1997})\ pp.\ \bibinfo {page} {176}\BibitemShut {NoStop}%
\bibitem [{\citenamefont {Bennett}\ and\ \citenamefont
  {di~Vincenzo}(2000)}]{bennett00}%
  \BibitemOpen
  \bibfield  {author} {\bibinfo {author} {\bibfnamefont {C.~H.}\ \bibnamefont
  {Bennett}}\ and\ \bibinfo {author} {\bibfnamefont {D.~P.}\ \bibnamefont
  {di~Vincenzo}},\ }\href {\doibase 10.1038/35005001} {\bibfield  {journal}
  {\bibinfo  {journal} {Nature}\ }\textbf {\bibinfo {volume} {404}},\ \bibinfo
  {pages} {247} (\bibinfo {year} {2000})}\BibitemShut {NoStop}%
\bibitem [{\citenamefont {Wiseman}\ and\ \citenamefont
  {Ralph}(2006)}]{wiseman06}%
  \BibitemOpen
  \bibfield  {author} {\bibinfo {author} {\bibfnamefont {H.~M.}\ \bibnamefont
  {Wiseman}}\ and\ \bibinfo {author} {\bibfnamefont {J.~F.}\ \bibnamefont
  {Ralph}},\ }\href {http://stacks.iop.org/1367-2630/8/i=6/a=090} {\bibfield
  {journal} {\bibinfo  {journal} {New J. Phys.}\ }\textbf {\bibinfo {volume}
  {8}},\ \bibinfo {pages} {90} (\bibinfo {year} {2006})}\BibitemShut {NoStop}%
\bibitem [{\citenamefont {Wiseman}\ and\ \citenamefont
  {Bouten}(2008)}]{wiseman08}%
  \BibitemOpen
  \bibfield  {author} {\bibinfo {author} {\bibfnamefont {H.~M.}\ \bibnamefont
  {Wiseman}}\ and\ \bibinfo {author} {\bibfnamefont {L.}~\bibnamefont
  {Bouten}},\ }\href {\doibase 10.1007/s11128-008-0075-8} {\bibfield  {journal}
  {\bibinfo  {journal} {Quantum Information Processing}\ }\textbf {\bibinfo
  {volume} {7}},\ \bibinfo {pages} {71} (\bibinfo {year} {2008})}\BibitemShut
  {NoStop}%
\bibitem [{\citenamefont {Combes}\ \emph {et~al.}(2008)\citenamefont {Combes},
  \citenamefont {Wiseman},\ and\ \citenamefont {Jacobs}}]{combes08}%
  \BibitemOpen
  \bibfield  {author} {\bibinfo {author} {\bibfnamefont {J.}~\bibnamefont
  {Combes}}, \bibinfo {author} {\bibfnamefont {H.~M.}\ \bibnamefont {Wiseman}},
  \ and\ \bibinfo {author} {\bibfnamefont {K.}~\bibnamefont {Jacobs}},\ }\href
  {\doibase 10.1103/PhysRevLett.100.160503} {\bibfield  {journal} {\bibinfo
  {journal} {Phys. Rev. Lett.}\ }\textbf {\bibinfo {volume} {100}},\ \bibinfo
  {eid} {160503} (\bibinfo {year} {2008})}\BibitemShut {NoStop}%
\bibitem [{\citenamefont {Combes}\ \emph {et~al.}(2010)\citenamefont {Combes},
  \citenamefont {Wiseman},\ and\ \citenamefont {Scott}}]{combes10}%
  \BibitemOpen
  \bibfield  {author} {\bibinfo {author} {\bibfnamefont {J.}~\bibnamefont
  {Combes}}, \bibinfo {author} {\bibfnamefont {H.~M.}\ \bibnamefont {Wiseman}},
  \ and\ \bibinfo {author} {\bibfnamefont {A.~J.}\ \bibnamefont {Scott}},\
  }\href {\doibase 10.1103/PhysRevA.81.020301} {\bibfield  {journal} {\bibinfo
  {journal} {Phys. Rev. A}\ }\textbf {\bibinfo {volume} {81}},\ \bibinfo
  {pages} {020301} (\bibinfo {year} {2010})}\BibitemShut {NoStop}%
\bibitem [{Note1()}]{Note1}%
  \BibitemOpen
  \bibinfo {note} {See Supplemental Material, which includes Refs.~\cite
  {gunter99,james01,white07} for additional derivations, calculation
details, asymptotic behaviour and experimental details}\BibitemShut {NoStop}%
\bibitem [{\citenamefont {Rudolph}(1999)}]{gunter99}%
  \BibitemOpen
  \bibfield  {author} {\bibinfo {author} {\bibfnamefont {G.}~\bibnamefont
  {Rudolph}},\ }
  \ \bibinfo {type} {Tech. Rep.}\ \bibinfo
  {number}{\href
  {http://ls11-www.informatik.uni-dortmund.de/people/rudolph/publications/papers/CI75.pdf} {CI-75/99}}\ , \bibinfo  {institution} {Department of Computer
  Science, University of Dortmund},\ \bibinfo {year} {1999}\BibitemShut
  {NoStop}%
\bibitem [{\citenamefont {James}\ \emph {et~al.}(2001)\citenamefont {James},
  \citenamefont {Kwiat}, \citenamefont {Munro},\ and\ \citenamefont
  {White}}]{james01}%
  \BibitemOpen
  \bibfield  {author} {\bibinfo {author} {\bibfnamefont {D.~F.~V.}\
  \bibnamefont {James}}, \bibinfo {author} {\bibfnamefont {P.~G.}\ \bibnamefont
  {Kwiat}}, \bibinfo {author} {\bibfnamefont {W.~J.}\ \bibnamefont {Munro}}, \
  and\ \bibinfo {author} {\bibfnamefont {A.~G.}\ \bibnamefont {White}},\ }\href
  {\doibase http://dx.doi.org/10.1103/PhysRevA.64.052312} {\bibfield  {journal}
  {\bibinfo  {journal} {Phys. Rev. A}\ }\textbf {\bibinfo {volume} {64}},\
  \bibinfo {pages} {052312} (\bibinfo {year} {2001})}\BibitemShut {NoStop}%
\bibitem [{\citenamefont {White}\ \emph {et~al.}(2007)\citenamefont {White},
  \citenamefont {Gilchrist}, \citenamefont {Pryde}, \citenamefont {O'Brien},
  \citenamefont {Bremner},\ and\ \citenamefont {Langford}}]{white07}%
  \BibitemOpen
  \bibfield  {author} {\bibinfo {author} {\bibfnamefont {A.~G.}\ \bibnamefont
  {White}}, \bibinfo {author} {\bibfnamefont {A.}~\bibnamefont {Gilchrist}},
  \bibinfo {author} {\bibfnamefont {G.~J.}\ \bibnamefont {Pryde}}, \bibinfo
  {author} {\bibfnamefont {J.~L.}\ \bibnamefont {O'Brien}}, \bibinfo {author}
  {\bibfnamefont {M.~J.}\ \bibnamefont {Bremner}}, \ and\ \bibinfo {author}
  {\bibfnamefont {N.~K.}\ \bibnamefont {Langford}},\ }\href {\doibase
  10.1364/JOSAB.24.000172} {\bibfield  {journal} {\bibinfo  {journal} {J. Opt.
  Soc. Am. B}\ }\textbf {\bibinfo {volume} {24}},\ \bibinfo {pages} {172}
  (\bibinfo {year} {2007})}\BibitemShut {NoStop}%
\end{thebibliography}
%

\end{document}